\documentclass[reprint,
twocolumn,
superscriptaddress,amsmath,amssymb,aps,prc,
showpacs,
showkeys,
floatfix]{revtex4}

\usepackage{graphicx}
\usepackage{hyperref}
\usepackage{url}
\usepackage{color} 
\usepackage[T1]{fontenc}
\usepackage{comment}
\usepackage{epsfig}
\usepackage{float}

\begin{document}

\title{Search for the $\boldsymbol{\eta}$ mesic $\boldsymbol{^{3}\hspace{-0.03cm} \textrm{He}}$ in the $\boldsymbol{pd\rightarrow d p \pi^{0}}$ reaction \\ with the WASA-at-COSY facility}

\date{\today}

\newcommand*{\IKPUU}{Division of Nuclear Physics, Department of Physics and Astronomy, Uppsala University, Box 516, 75120 Uppsala, Sweden}
\newcommand*{\ASWarsN}{Department of Nuclear Physics, National Centre for Nuclear Research, Ludwika Pasteura~7, 02-093, Warsaw, Poland}
\newcommand*{\IPJ}{Institute of Physics, Jagiellonian University, prof.\ Stanis{\l}awa {\L}ojasiewicza~11, 30-348 Krak\'{o}w, Poland}
\newcommand*{\Kitz}{Kitzb\"uhel Centre for Physics, Kitzb\"uhel, Austria}
\newcommand*{\PITue}{Physikalisches Institut, Eberhard--Karls--Universit\"at 
	T\"ubingen, Auf der Morgenstelle~14, 72076 T\"ubingen, Germany}
\newcommand*{\Edinb}{School of Physics and Astronomy, University of Edinburgh, James Clerk Maxwell Building, Peter Guthrie Tait Road, Edinburgh EH9 3FD, Great Britain}
\newcommand*{\MS}{Institut f\"ur Kernphysik, Westf\"alische Wilhelms--Universit\"at M\"unster, Wilhelm--Klemm--Str.~9, 48149 M\"unster, Germany}
\newcommand*{\PGI}{Peter Gr\"unberg Institut, PGI--6 Elektronische Eigenschaften, Forschungszentrum J\"ulich, 52425 J\"ulich, Germany}
\newcommand*{\DUS}{Institut f\"ur Laser-- und Plasmaphysik, Heinrich--Heine Universit\"at D\"usseldorf, Universit\"atsstr.~1, 40225 D\"usseldorf, Germany}
\newcommand*{\ASWarsH}{High Energy Physics Department, National Centre for Nuclear Research, Ludwika Pasteura~7, 02-093, Warsaw, Poland}
\newcommand*{\Budker}{Budker Institute of Nuclear Physics of SB RAS, 11~akademika Lavrentieva prospect, Novosibirsk, 630090, Russia}
\newcommand*{\Novosib}{Novosibirsk State University, 2~Pirogova Str., Novosibirsk, 630090, Russia}
\newcommand*{\IITB}{Department of Physics, Indian Institute of Technology Bombay, Powai, Mumbai--400076, Maharashtra, India}
\newcommand*{\IKPJ}{Institut f\"ur Kernphysik, Forschungszentrum J\"ulich, 52425 J\"ulich, Germany}
\newcommand*{\NWU}{Department of Physics, Nara Women's University, Nara 630-8506, Japan}
\newcommand*{\ANDES}{Departamento de Fisica, Universidad de los Andes, Cra.~1E, 18A--10, Bogot{\'a}, Colombia}
\newcommand*{\HeJINR}{Veksler and Baldin Laboratory of High Energiy Physics, Joint Institute for Nuclear Physics, 6~Joliot--Curie, Dubna, 141980, Russia}
\newcommand*{\Bochum}{Institut f\"ur Experimentalphysik I, Ruhr--Universit\"at Bochum, Universit\"atsstr.~150, 44780 Bochum, Germany}
\newcommand*{\ZELJ}{Zentralinstitut f\"ur Elektronik, Forschungszentrum J\"ulich, 52425 J\"ulich, Germany}
\newcommand*{\Erl}{Physikalisches Institut, Friedrich--Alexander--Universit\"at Erlangen--N\"urnberg, Erwin--Rommel-Str.~1, 91058 Erlangen, Germany}
\newcommand*{\ITEP}{Institute for Theoretical and Experimental Physics named by A.I.\ Alikhanov of National Research Centre ``Kurchatov Institute'', 25~Bolshaya Cheremushkinskaya, Moscow, 117218, Russia}
\newcommand*{\Giess}{II.\ Physikalisches Institut, Justus--Liebig--Universit\"at Gie{\ss}en, Heinrich--Buff--Ring~16, 35392 Giessen, Germany}
\newcommand*{\IITI}{Department of Physics, Indian Institute of Technology Indore, Khandwa Road, Indore--452017, Madhya Pradesh, India}
\newcommand*{\HepGat}{High Energy Physics Division, Petersburg Nuclear Physics Institute named by B.P.\ Konstantinov of National Research Centre ``Kurchatov Institute'', 1~mkr.\ Orlova roshcha, Leningradskaya Oblast, Gatchina, 188300, Russia}
\newcommand*{\Katow}{August Che{\l}kowski Institute of Physics, University of Silesia, Uniwersytecka~4, 40-007, Katowice, Poland}
\newcommand*{\IFJ}{The Henryk Niewodnicza{\'n}ski Institute of Nuclear Physics, Polish Academy of Sciences, 152~Radzikowskiego St, 31-342 Krak\'{o}w, Poland}
\newcommand*{\KEK}{High Energy Accelerator Research Organisation KEK, Tsukuba, Ibaraki 305--0801, Japan}
\newcommand*{\ASLodz}{Department of Cosmic Ray Physics, National Centre for  Nuclear Research, ul.\ Uniwersytecka~5, 90--950 {\L}\'{o}d\'{z}, Poland}
\newcommand*{\Kepler}{Kepler Center f\"ur Astro-- und Teilchenphysik, Physikalisches Institut der Universit\"at T\"ubingen, Auf der Morgenstelle~14, 72076 T\"ubingen, Germany}
\newcommand*{\NCBJ}{High Energy Physics Division, National Centre for Nuclear Research, 05-400 Otwock-Świerk, Poland}
\newcommand*{\NITJ}{Department of Physics, Malaviya National Institute of Technology Jaipur, JLN Marg Jaipur - 302017, Rajasthan, India}
\newcommand*{\JARA}{JARA--FAME, J\"ulich Aachen Research Alliance, Forschungszentrum J\"ulich, 52425 J\"ulich, and RWTH Aachen, 52056 Aachen, Germany}.
\newcommand*{\Tomsk}{Department of Physics, Tomsk State University, 36~Lenina Avenue, Tomsk, 634050, Russia}

\author{P.~Adlarson}    \affiliation{\IKPUU}
\author{W.~Augustyniak} \affiliation{\ASWarsN}
\author{M.~Bashkanov}   \affiliation{\Edinb}
\author{S.~D.~Bass}		\affiliation{\IPJ}\affiliation{\Kitz}
\author{F.S.~Bergmann}  \affiliation{\MS}
\author{M.~Ber{\l}owski}\affiliation{\ASWarsH}
\author{A.~Bondar}		\affiliation{\Budker}\affiliation{\Novosib}
\author{M.~B\"uscher}   \affiliation{\PGI}\affiliation{\DUS}
\author{H.~Cal\'{e}n}   \affiliation{\IKPUU}
\author{I.~Ciepa{\l}}   \affiliation{\IFJ}
\author{H.~Clement}     \affiliation{\PITue}\affiliation{\Kepler}
\author{E.~Czerwi{\'n}ski}\affiliation{\IPJ}
\author{K.~Demmich}     \affiliation{\MS}
\author{R.~Engels}      \affiliation{\IKPJ}
\author{A.~Erven}       \affiliation{\ZELJ}
\author{W.~Erven}       \affiliation{\ZELJ}
\author{W.~Eyrich}      \affiliation{\Erl}
\author{P.~Fedorets}	\affiliation{\IKPJ}\affiliation{\ITEP}
\author{K.~F\"ohl}      \affiliation{\Giess}
\author{K.~Fransson}    \affiliation{\IKPUU}
\author{F.~Goldenbaum}  \affiliation{\IKPJ}
\author{A.~Goswami}     \affiliation{\IITI}\affiliation{\IKPJ}
\author{K.~Grigoryev}	\affiliation{\IKPJ}\affiliation{\HepGat}
\author{L.~Heijkenskj\"old}	\altaffiliation[Current address: ]{\Mainz}\affiliation{\IKPUU}
\author{V.~Hejny}       \affiliation{\IKPJ}
\author{N. H\"usken}	\affiliation{\MS}
\author{S.~Hirenzaki}	\affiliation{\NWU}
\author{T.~Johansson}   \affiliation{\IKPUU}
\author{B.~Kamys}       \affiliation{\IPJ}
\author{N.~G.~Kelkar}	\affiliation{\ANDES}
\author{G.~Kemmerling}  \altaffiliation[Current address: ]{\JCNS}\affiliation{\ZELJ}
\author{A.~Khoukaz}     \affiliation{\MS}
\author{A.~Khreptak}	\affiliation{\IPJ}
\author{D.~A.~Kirillov}	\affiliation{\HeJINR}
\author{S.~Kistryn}     \affiliation{\IPJ}
\author{H.~Kleines}     \altaffiliation[Current address: ]{\JCNS}\affiliation{\ZELJ}
\author{B.~K{\l}os}     \affiliation{\Katow}
\author{W.~Krzemie{\'n}}	\affiliation{\NCBJ}
\author{P.~Kulessa}     \affiliation{\IFJ}
\author{A.~Kup\'{s}\'{c}}	\affiliation{\IKPUU}\affiliation{\ASWarsH}
\author{K.~Lalwani}     \affiliation{\NITJ}
\author{D.~Lersch}      \altaffiliation[Current address: ]{\Florida}\affiliation{\IKPJ}
\author{B.~Lorentz}     \affiliation{\IKPJ}
\author{A.~Magiera}     \affiliation{\IPJ}
\author{R.~Maier}       \affiliation{\IKPJ}\affiliation{\JARA}
\author{P.~Marciniewski}\affiliation{\IKPUU}
\author{B.~Maria{\'n}ski}\affiliation{\ASWarsN}
\author{H.--P.~Morsch}  \affiliation{\ASWarsN}
\author{P.~Moskal}      \affiliation{\IPJ}
\author{H.~Ohm}			\affiliation{\IKPJ}
\author{W.~Parol}		\affiliation{\IPJ}
\author{E.~Perez del Rio}	\affiliation{\PITue}\affiliation{\Kepler}
\author{N.~M.~Piskunov}	\affiliation{\HeJINR}
\author{D.~Prasuhn}     \affiliation{\IKPJ}
\author{D.~Pszczel}     \affiliation{\IKPUU}\affiliation{\ASWarsH}
\author{K.~Pysz}        \affiliation{\IFJ}
\author{J.~Ritman}\affiliation{\IKPJ}\affiliation{\JARA}\affiliation{\Bochum}
\author{A.~Roy}			\affiliation{\IITI}
\author{O.~Rundel}		\affiliation{\IPJ}
\author{S.~Sawant}      \affiliation{\IITB}
\author{S.~Schadmand}   \affiliation{\IKPJ}
\author{T.~Sefzick}     \affiliation{\IKPJ}
\author{V.~Serdyuk}		\affiliation{\IKPJ}
\author{B.~Shwartz}		\affiliation{\Budker}\affiliation{\Novosib}
\author{T.~Skorodko}    \affiliation{\PITue}\affiliation{\Kepler}\affiliation{\Tomsk}
\author{M.~Skurzok}     \altaffiliation[Current address: ]{\INFN}\affiliation{\IPJ}
\author{J.~Smyrski}     \affiliation{\IPJ}
\author{V.~Sopov}       \affiliation{\ITEP}
\author{R.~Stassen}     \affiliation{\IKPJ}
\author{J.~Stepaniak}   \affiliation{\ASWarsH}
\author{E.~Stephan}     \affiliation{\Katow}
\author{G.~Sterzenbach} \affiliation{\IKPJ}
\author{H.~Stockhorst}  \affiliation{\IKPJ}
\author{H.~Str\"oher}   \affiliation{\IKPJ}\affiliation{\JARA}
\author{A.~Szczurek}    \affiliation{\IFJ}
\author{M.~Wolke}       \affiliation{\IKPUU}
\author{A.~Wro{\'n}ska} \affiliation{\IPJ}
\author{P.~W\"ustner}   \affiliation{\ZELJ}
\author{A.~Yamamoto}    \affiliation{\KEK}
\author{J.~Zabierowski} \affiliation{\ASLodz}
\author{M.J.~Zieli{\'n}ski}	\affiliation{\IPJ}
\author{J.~Z{\l}oma{\'n}czuk}\affiliation{\IKPUU}
\author{M.~{\.Z}urek}   \altaffiliation[Current address: ]{\Berkeley}\affiliation{\IKPJ}

\newcommand*{\JCNS}{J\"ulich Centre for Neutron Science JCNS, Forschungszentrum J\"ulich, 52425 J\"ulich, Germany}
\newcommand*{\INFN}{INFN, Laboratori Nazionali di Frascati, Via E. Fermi, 40, 00044 Frascati (Roma), Italy}
\newcommand*{\Berkeley}{Lawrence Berkeley National Laboratory, Berkeley, California 94720}
\newcommand*{\Mainz}{Institut f\"ur Kernphysik, Johannes Guten\-berg--Universit\"at Mainz, Johann--Joachim--Becher Weg~45, 55128 Mainz, Germany}
\newcommand*{\Florida}{Department of Physics, Florida State University, 77 Chieftan Way, Tallahassee, FL 32306-4350, USA} 

\collaboration{WASA-at-COSY Collaboration}\noaffiliation

\begin{abstract}
\noindent
The excitation function for the $pd\rightarrow d p \pi^{0}$ reaction
has been measured by WASA-at-COSY experiment with the aim of searching for $^{3}\hspace{-0.03cm}\textrm{He-}\eta$ mesic nuclei. The measurement in the vicinity of $\eta$ meson production was performed using a ramped proton beam. The data analysis and interpretation was carried out with the assumption that the $\eta$-mesic Helium decays via the formation of an intermediate 
$\textrm{N}^{\ast}$(1535) resonance. 
No direct signal of the $\eta$-mesic nucleus is observed in the excitation function.
We determine a new improved upper limit for the total cross section for the bound state production and decay in the
$pd\rightarrow (^{3}\hspace{-0.03cm}\textrm{He-}\eta)_{bound} \rightarrow d p \pi^{0}$ process. 
It varies between 13~nb to 24~nb for the bound state with width in the range $\Gamma \in (5,50)$~MeV.

\end{abstract}


\pacs{21.85.+d, 21.65.Jk, 25.80.-e, 13.75.-n}
\keywords{mesic nuclei, $\eta$-mesic nucleus, $\eta$ meson}

\maketitle

\section{Introduction}

In this paper
we present a new high statistics search for $^{3}\hspace{-0.03cm}\textrm{He-}\eta$ bound states 
with focus on the $pd \rightarrow d p \pi^0$ reaction.
The measurement was performed 
using data from the WASA-at-COSY experiment at Forschungszentrum J\"ulich.
Strong attractive interactions between
the $\eta$ meson and nucleons mean that
there is a chance
to form $\eta$ meson bound states in nuclei \cite{Haider:1986sa}.
If discovered in experiments, 
these mesic nuclei would be a new state of 
matter bound just by the strong interaction
without electromagnetic Coulomb effects playing a role because of 
the zero electric charge of the $\eta$ meson.
Early experiments with low statistics
using photon~\cite{Pheron:2012aj,Baskov:2012yd}, 
pion~\cite{Chrien:1988gn}, 
proton~\cite{Budzanowski:2008fr} 
or deuteron~\cite{Afanasiev:2011zza,Moskal:2010ee,AdlarsonPRC2013}
beams 
gave hints for possible $\eta$ mesic 
bound states but no clear signal \cite{Kelkar:2013lwa,Metag:2017yuh}.
The new results reported here 
are complementary to the 
recent
$^{3}\hspace{-0.03cm}\textrm{He-}\eta$
bound state search using the
$p d \rightarrow {^{3}\hspace{-0.03cm}\textrm{He}} 2 \gamma$
and 
$p d \rightarrow {^{3}\hspace{-0.03cm}\textrm{He}} 6 \gamma$
reactions
and 
performed with the same experiment.

The key physical process 
involves a virtual $\eta$ meson
produced in the $pd$ collision
forming a bound state with the 
$^{3}\hspace{-0.03cm}\textrm{He}$ nucleus in which it
is produced.
The bound states might form by the attractive interaction, with finite 
width corresponding to the finite lifetime of the state due to the absorptive interaction with the nucleus. 
$\eta$ meson interactions with nucleons and nuclei are a topic 
of much experimental and 
theoretical interest.
For recent reviews see 
\cite{Kelkar:2013lwa,Metag:2017yuh,Bass:2018xmz,Krusche:2014ava,Wilkin:2016mfn,SkurzokFBS2020}.

Hints for possible $\eta$ helium bound states are inferred from the observation of strong interaction in the $\eta$ helium system. One finds a sharp rise in the cross section 
at threshold for $\eta$ production in
photoproduction from $^{3}\hspace{-0.03cm}\textrm{He}$~\cite{Pheron:2012aj,Pfeiffer:2003zd}
and in the proton-deuteron reaction
$dp\rightarrow {^{3}\hspace{-0.03cm}\textrm{He}} \eta$~\cite{Adlarson:2018rgs}.
These observations may hint at a reduced $\eta$ effective mass in the nuclear medium,
see e.g.
\cite{Bass:2018xmz}.

Possible $\eta$-nucleus binding energies are related
to the $\eta$-nucleus optical potential and to the
value of $\eta$-nucleon scattering length $a_{\eta N}$
\cite{Ericson:1988gk}. 
Phenomenological 
estimates for the real part 
of $a_{\eta N}$ are typically
between 0.2 and 1 fm.
$\eta$ bound states in helium require a large $\eta$-nucleon
scattering length 
with real part greater than about 
0.7--1.1~fm  
\cite{Barnea:2017epo,Barnea:2017oyk,Fix:2017ani}.
Recent calculations in the framework of 
optical potential~\cite{Xie:2016zhs},
multi-body calculations~\cite{Barnea:2017oyk},
and pionless effective field theory~\cite{Barnea:2017epo}
suggest a possible $^{3}\hspace{-0.03cm}\textrm{He-}\eta$ bound state.

The related system of $\eta'$-nucleus interactions is also a
strong candidate for a meson-nucleus
bound state. 
Recent measurements 
by the CBELSA/TAPS collaboration 
in Bonn 
using photoproduction of $\eta'$
mesons from a carbon target
determined the $\eta'$-nucleus optical
potential
$V_{\rm opt} = V + iW$
with
the strength of the real part
at nuclear matter density
$\rho_0$
related to the meson's effective mass shift
$V 
= m^{\ast} - m 
= -37 \pm 10 \pm 10 \ {\rm MeV}
$
and imaginary part
$
W
= -10 \pm 2.5 \ {\rm MeV}$
at 
$\rho_0$
\cite{Nanova:2013fxl}.
With the attractive
real part of the potential
greater than the imaginary part,
this result has inspired a 
program of bound state searches
with first results 
(ruling out much larger potential
 depths)
reported in Ref. \cite{Tanaka} 
and future more accurate measurements in planning.
The $\eta'$ mass shift suggested
by CBELSA/TAPS is very close to the prediction of the
Quark Meson Coupling model, QMC,
with mixing angle -20 degrees \cite{Bass:2005hn,Bass:2013nya}
and consistent with
$\eta'$-nucleon
scattering length
determinations from
Bonn \cite{Anisovich:2018yoo}
and 
COSY-11 \cite{Czerwinski:2014yot}.
The QMC model
predicts an $\eta$ nucleus potential depth about -100 MeV 
at 
$\rho_0$.

\begin{figure*}[ht!]
	\centering
	\includegraphics[width=0.78\textwidth]{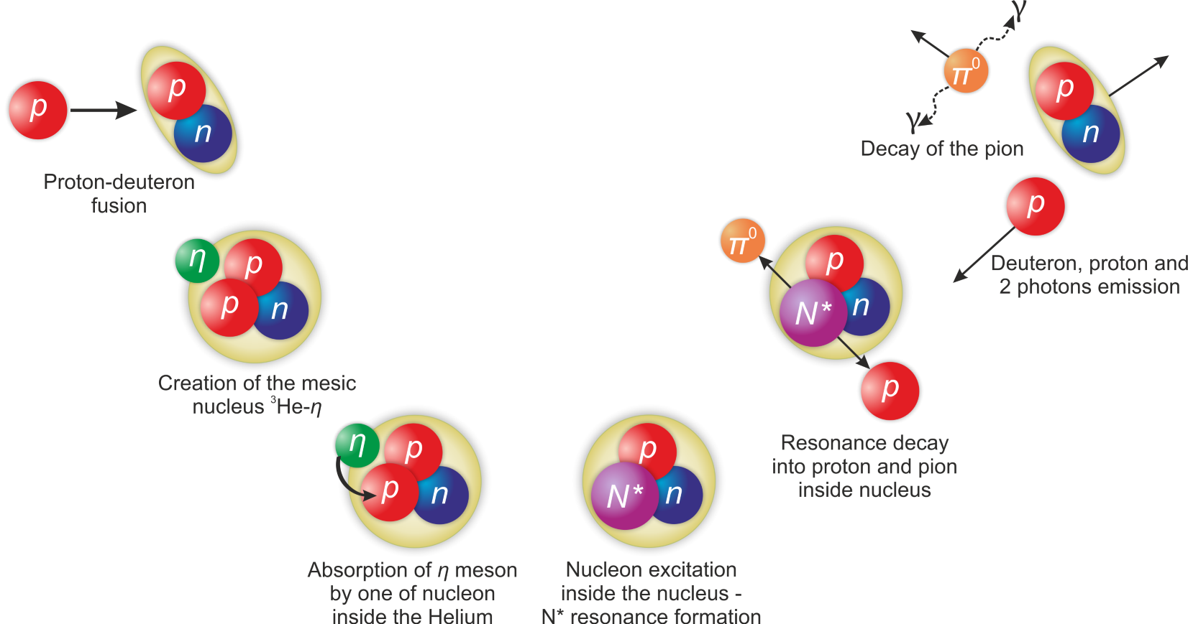}
	\caption{(Color online) Model of the $^{3}\hspace{-0.03cm}\textrm{He-}\eta$ bound state production and decay in the $pd \rightarrow d p \pi^{0}$ reaction.}\label{fig_model}
\end{figure*}

In May 2014 the experimental search for $\eta$ mesic $^{3}\hspace{-0.03cm}\textrm{He}$ nuclei was carried out using the WASA-at-COSY detection system~\cite{COSY_description1,COSY_description5,WASA_description1,WASA_description2,WASA_description3,WASA_pellett_target} at Forschungszentrum J\"ulich in Germany 
colliding the COSY proton beam with a deuteron pellet target. The search for $\eta$-mesic bound states was performed considering two main predicted mechanisms for the $\eta$-mesic bound state decay, via the formation of an intermediate $\textrm{N}^{\ast}$(1535) resonance and its decay into a nucleon pion pair 
(used in previous experimental studies) 
and via decay of $\eta$-meson still ``orbiting'' around the nucleus 
\cite{SkurzokNPA2020}. 
The bound state, if it exists, would be manifest as a resonance structure in the excitation function for the studied processes below the $pd \rightarrow {^{3}\hspace{-0.03cm}\textrm{He}}\eta$ reaction threshold.

The mechanism of $\eta$-mesic $^{3}\hspace{-0.03cm}\textrm{He}$ decay has been investigated recently for the first time by analysing the $pd \rightarrow {^{3}\hspace{-0.03cm}\textrm{He}} 2 \gamma$ and $pd \rightarrow {^{3}\hspace{-0.03cm}\textrm{He}} 6 \gamma$ reactions~\cite{AdlarsonPLB2020} assuming the
theoretical model
recently developed 
in~\cite{SkurzokNPA2020}. 
The final excitation functions for both channels showed a slight indication of the signal from a possible
bound state for $\Gamma > 20$~MeV and binding energies in the range from 0 to 15~MeV which is, however, covered by the systematic error. 
Therefore, drawing conclusions for the bound state existence in the considered mechanism was not possible.
The upper limit at the CL=90\% obtained by fitting simultaneously excitation functions for both processes varied between 2 nb to 15 nb depending on the bound state parameters~\cite{AdlarsonPLB2020}.

In this paper we present results of the search for $\eta$-mesic $^{3}\hspace{-0.03cm}\textrm{He}$ in the $pd \rightarrow d p \pi^{0}$ reaction corresponding to the mechanism 
$pd\rightarrow (^{3}\hspace{-0.03cm}\textrm{He-}\eta)_{bound} \rightarrow \textrm{N}^{\ast}d \rightarrow d p \pi^{0}$ 
via excitation of the $\textrm{N}^{\ast}$(1535)
resonance
-- see Fig.~1 --
with the $\textrm{N}^{\ast}$(1535) coming with 
narrower momentum distribution 
compared to nucleons
\cite{Kelkar:2019hjm,Kelkar:2020wmh}.

Earlier bound state searches at COSY, assuming the above mechanism, focused on the reaction
$dd \rightarrow {^{3}\hspace{-0.03cm}\textrm{HeN}} \pi$. 
The excitation functions determined around the threshold for 
$dd \rightarrow {^{4}\hspace{-0.03cm}\textrm{He}} \eta$
did not reveal a structure 
that could be interpreted as a narrow mesic nucleus~\cite{AdlarsonPRC2013,Adlarson:2016dme,Skurzok_PLB2018,Skurzok_APP2020}. 
Upper limits for the total cross sections for bound state production and decay in the processes
$dd \rightarrow (^{4}\hspace{-0.03cm}\textrm{He-}\eta)_{bound} \rightarrow {^{3}\hspace{-0.03cm}\textrm{He}} n \pi^{0}$ 
and 
$dd \rightarrow (^{4}\hspace{-0.03cm}\textrm{He-}\eta)_{bound} \rightarrow {^{3}\hspace{-0.03cm}\textrm{He}} p \pi^{-}$ 
were deduced to be about 5~nb and 10~nb for 
the $n \pi^{0}$ and $p \pi^{-}$ channels, respectively~\cite{Adlarson:2016dme}.
The bound state production 
cross sections 
for $pd \rightarrow (^{3}\hspace{-0.03cm}\textrm{He-}\eta)_{bound}$ \cite{Wilkin:2014mla} 
are expected to be more than 20 times larger than for
$dd \rightarrow (^{4}\hspace{-0.03cm}\textrm{He-}\eta)_{bound}$~\cite{Wycech:2014wua}.


\section{Experiment}

\subsection{Measurement conditions}

The high statistics experiment devoted to the search for $^{3}\hspace{-0.03cm}\textrm{He-}\eta$ mesic nuclei in the $pd \rightarrow d p \pi^{0}$ reaction was carried out with the WASA (Wide Angle Shower Apparatus)~\cite{Adlarson:2018rgs,WASA_description1,WASA_description2,WASA_description3,WASA_pellett_target} detection setup installed at the COSY accelerator~\cite{COSY_description1,COSY_description5}. 
The WASA detector consisted of two main parts: the Forward Detector (FD) and Central Detector (CD) optimized for tagging the recoil particles and registering the meson decay products, respectively.

The measurement was performed changing the proton beam momentum very slowly and continuously around the $\eta$ production threshold in each acceleration cycle from 1.426 to 1.635 GeV/c,  corresponding to the $^{3}\hspace{-0.03cm}\textrm{He}\eta$ excess energy range Q$\in$(-70,30)~MeV (Q=$\sqrt{s_{pd}} - m_{\eta} - m_{^{3}\hspace{-0.03cm}\textrm{He}}$, where $\sqrt{s_{pd}}$ is invariant mass of colliding proton and deuteron). The application of this so-called ramped beam technique allowed us to reduce the systematic uncertainties with respect to separate runs at fixed beam energies~\cite{AdlarsonPRC2013,SmyrskiPLB2007}. 

Possible resonance-like structure below the $\eta$ production threshold associated with the $^{3}\hspace{-0.03cm}\textrm{He-}\eta$ bound state was searched for via  measurement of the excitation function for the $pd \rightarrow d p \pi^{0}$ reaction. 


\subsection{$\boldsymbol{pd \rightarrow (^{3}\hspace{-0.03cm}\textrm{He-}\eta)_{bound} \rightarrow d p \pi^{0}}$ events selection}

The events corresponding to formation of $^{3}\hspace{-0.03cm}\textrm{He-}\eta$ bound states were selected with appropriate conditions based on the Monte Carlo simulation of the $pd \rightarrow (^{3}\hspace{-0.03cm}\textrm{He-}\eta)_{bound} \rightarrow d p \pi^{0}$ reaction. 
The considered kinematic mechanism of the process is presented schematically in Fig.~\ref{fig_model}. 
According to the scheme, the proton deuteron collision leads to the formation of a $^{3}\hspace{-0.03cm}\textrm{He}$ nucleus bound with the $\eta$ meson via strong interactions. Then, the $\eta$ meson can 
be absorbed by one of the nucleons inside the helium exciting it to the $\textrm{N}^{\ast}$(1535) nucleon resonance until the resonance decays into a
proton $\pi^{0}$ pair, with the pion
subsequently decaying into two photons.
This mechanism, with formation of an intermediate $\textrm{N}^{\ast}$, was also
assumed in the previous analyses~\cite{AdlarsonPRC2013,Adlarson:2016dme,Skurzok_PLB2018,Skurzok_APP2020}.

\begin{figure}[h!]
	\centering
	\includegraphics[width=0.99\columnwidth]{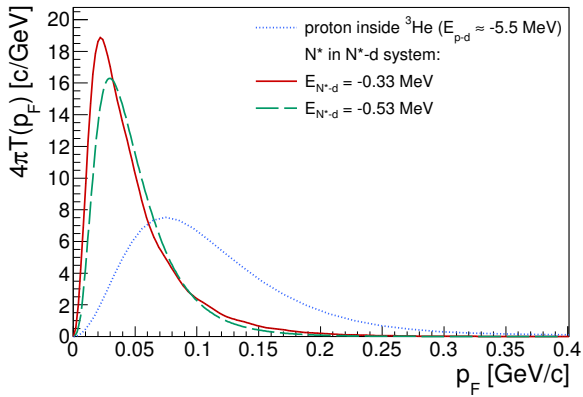}
	\caption{(Color online) Fermi momentum distribution of the $\textrm{N}^{\ast}$ resonance in the $\textrm{N}^{\ast}$-deuteron bound state for two different values of binding energy $E_{\textrm{N}^{\ast}\textrm{-}d} = -0.33$~MeV and $-0.53$~MeV (red solid and green dashed lines, respectively)~\cite{Kelkar:2019hjm,Kelkar:2020wmh} and of protons inside $^{3}\hspace{-0.03cm}\textrm{He}$ nucleus for the separation energy $\approx~5.5$~MeV (blue dotted line)~\cite{Nogga:PRC2003}.}
	\label{distr}
\end{figure}

The simulation was performed using the $\textrm{N}^{\ast}$ resonance momentum distribution in the $\textrm{N}^{\ast}$-deuteron system determined recently by Kelkar et al.~\cite{Kelkar:2019hjm,Kelkar:2020wmh}. 
The distribution calculated for two different values of binding energy $E_{\textrm{N}^{\ast}\textrm{-}d} = -0.33$~MeV and $-0.53$~MeV is shown in Fig.~\ref{distr} (red solid and green dashed lines). 
It is much narrower compared to the Fermi momentum distribution of protons inside $^{3}\hspace{-0.03cm}\textrm{He}$~\cite{Nogga:PRC2003} (blue dotted line) 
which results from the fact that the $\textrm{N}^{\ast}$ binding energy is smaller
than the energy separation of proton in $^{3}\hspace{-0.03cm}\textrm{He}$.

The deuteron in this process plays the role of a spectator. 
In the simulations it was assumed that the bound state has a resonance structure given by the Breit-Wigner distribution with fixed binding energy $B_s$ and width $\Gamma$:
\begin{equation}
    N(\sqrt{s_{pd}})= \frac{\Gamma\hspace{-0.03cm}^{2}/4}{\left(\sqrt{s_{pd}} - (m_{\eta} + m_{^{3}\hspace{-0.03cm}\textrm{He}} - B_{s})\right)^{2} + \Gamma\hspace{-0.03cm}^{2}/4},
\label{eq:BW_formula}
\end{equation}

\noindent
where $\sqrt{s_{pd}}$ is the invariant mass of the colliding proton and deuteron and $m_{\eta} + m_{^{3}\hspace{-0.03cm}\textrm{He}} - B_{s}$ is the bound state mass. 
The total invariant mass $\sqrt{s_{pd}}$ was calculated based on the proton beam momentum $p_{beam}$, which was generated with uniform probability density distribution in the range of $p_{beam} \in (1.426,1.635)$~GeV/c corresponding to the experimental beam ramping.

\begin{figure}[hb!]
	\centering
	\includegraphics[width=0.99\columnwidth]{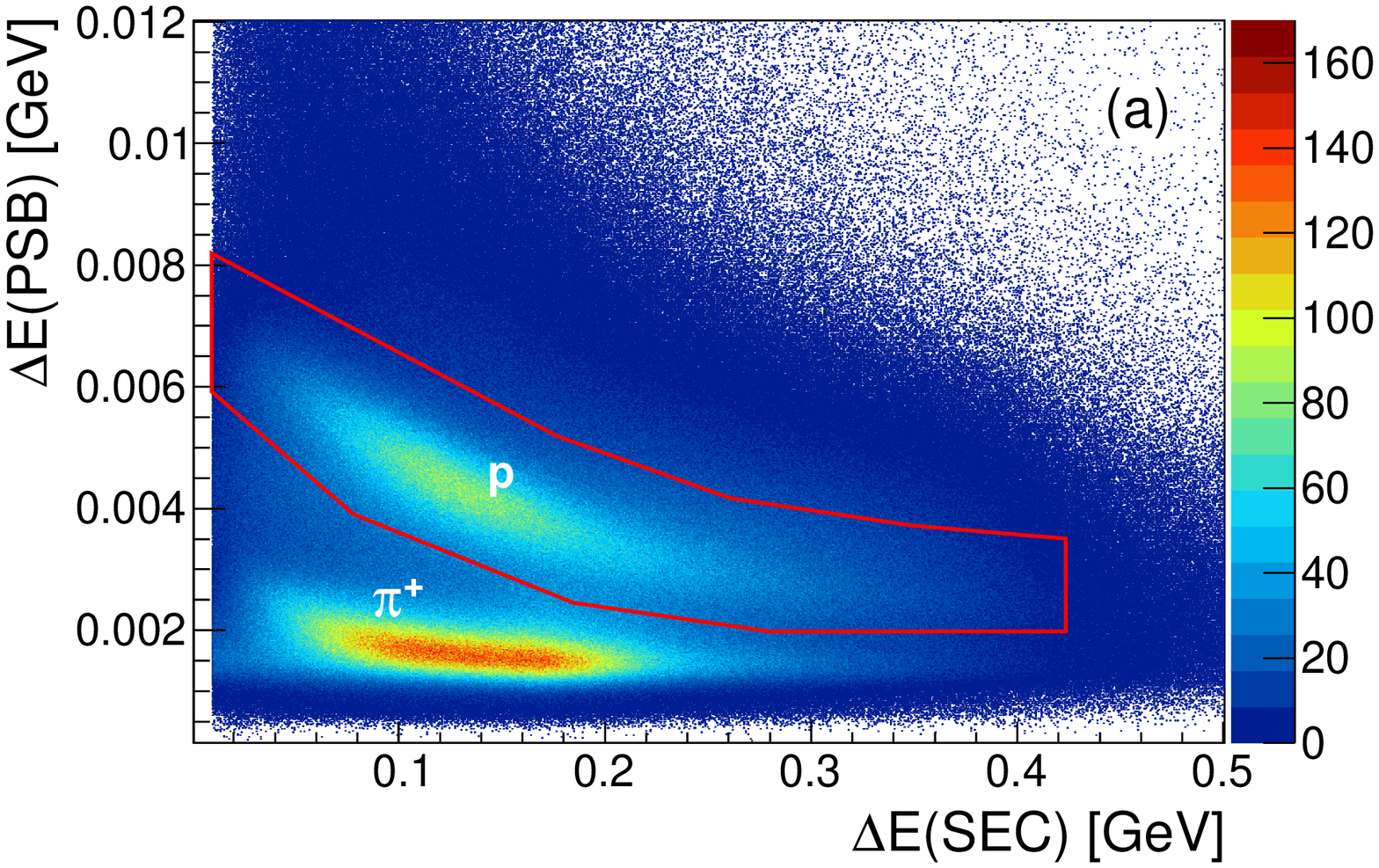}
	\includegraphics[width=0.99\columnwidth]{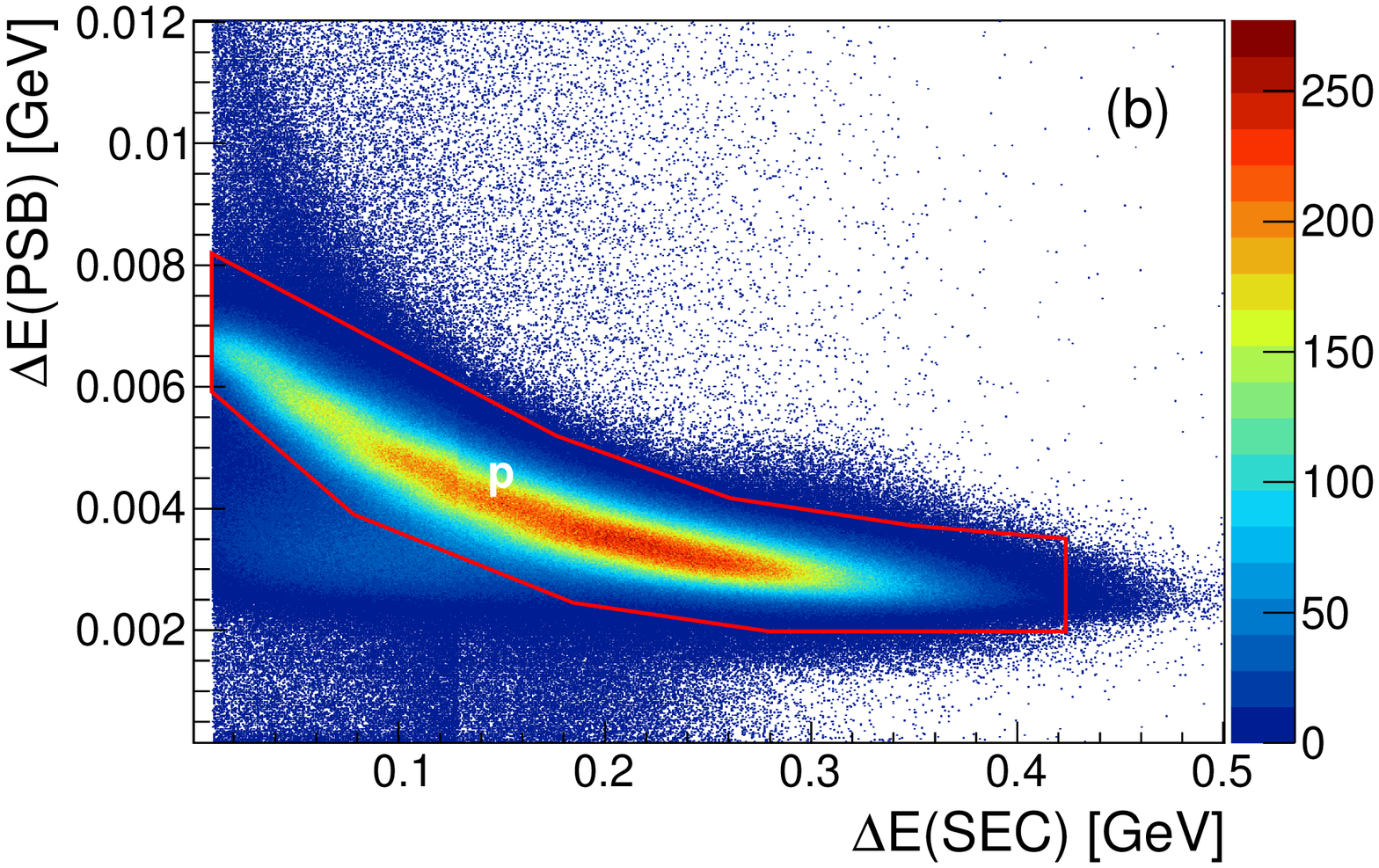}
	\caption{(Color online) Energy deposited in the Scintillator Electromagnetic Calorimeter (SEC) as a function of the energy loss in the Plastic Scintillator Barrel (PSB) for experimental data (a) and simulations (b). 
	The area corresponding to selected protons is marked with a red solid line.}
	\label{proton_cd}
\end{figure}

Events selection for the $pd \rightarrow (^{3}\hspace{-0.03cm}\textrm{He-}\eta)_{bound} \rightarrow d p \pi^{0}$ process started with particles identification in the Central Detector. 
Protons were identified based on the energy deposited in the Scintillator Electromagnetic Calorimeter (SEC) combined with the energy loss in the Plastic Scintillator Barrel (PSB),
see Fig.~\ref{proton_cd}. 

\begin{figure*}[h!]
    \centering
    \includegraphics[width=0.99\columnwidth]{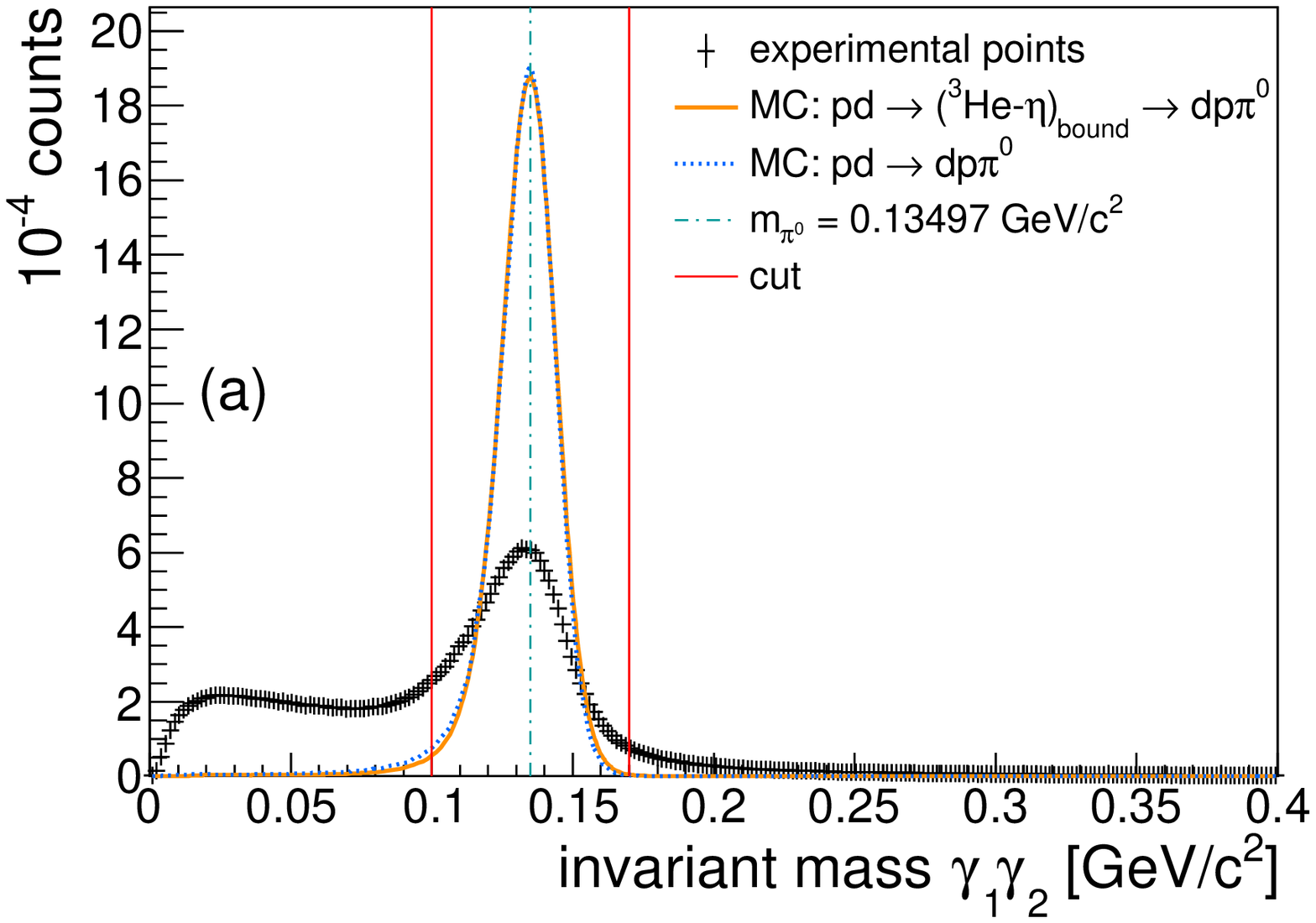}
	\includegraphics[width=0.99\columnwidth]{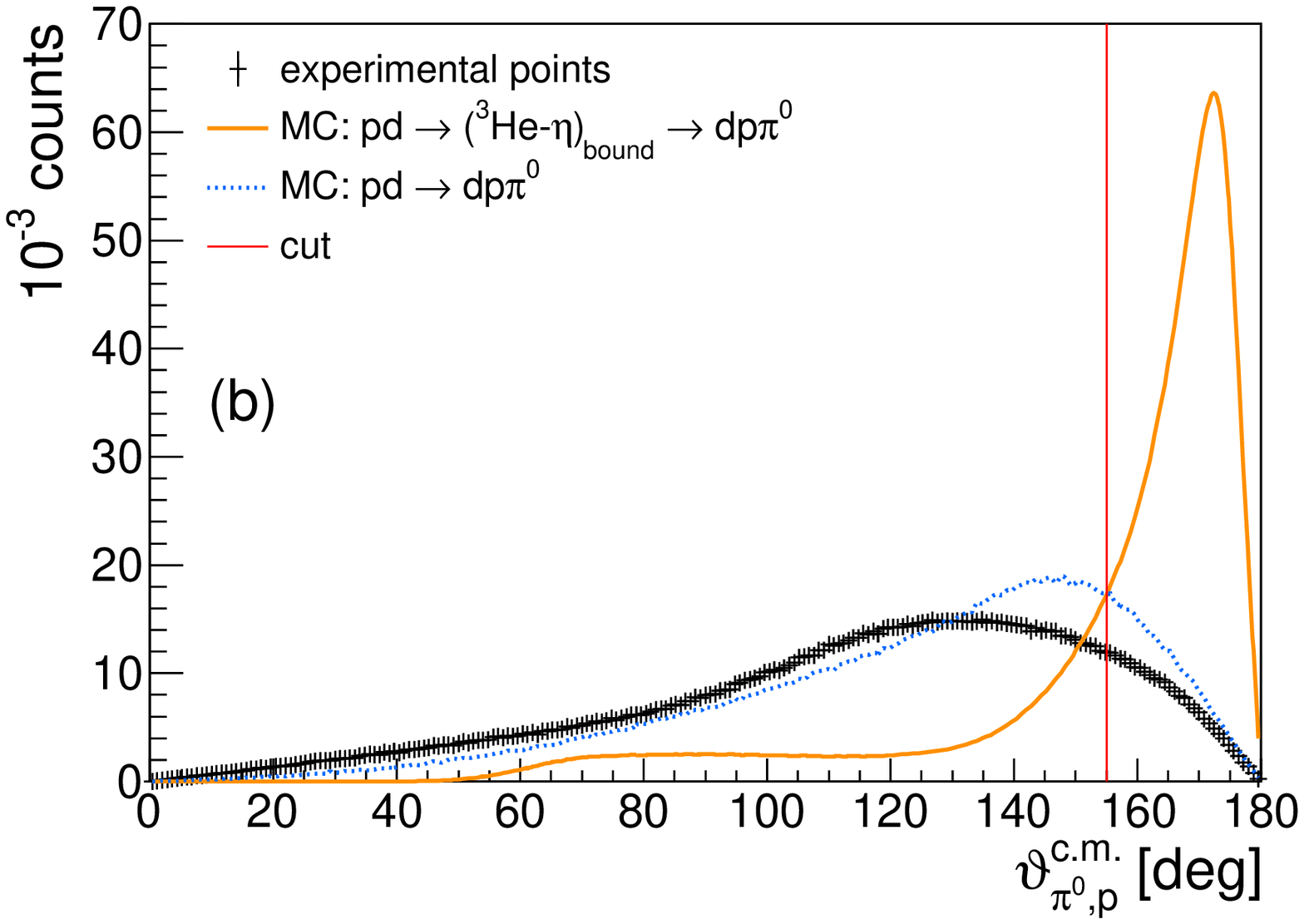}\\
	\includegraphics[width=0.99\columnwidth]{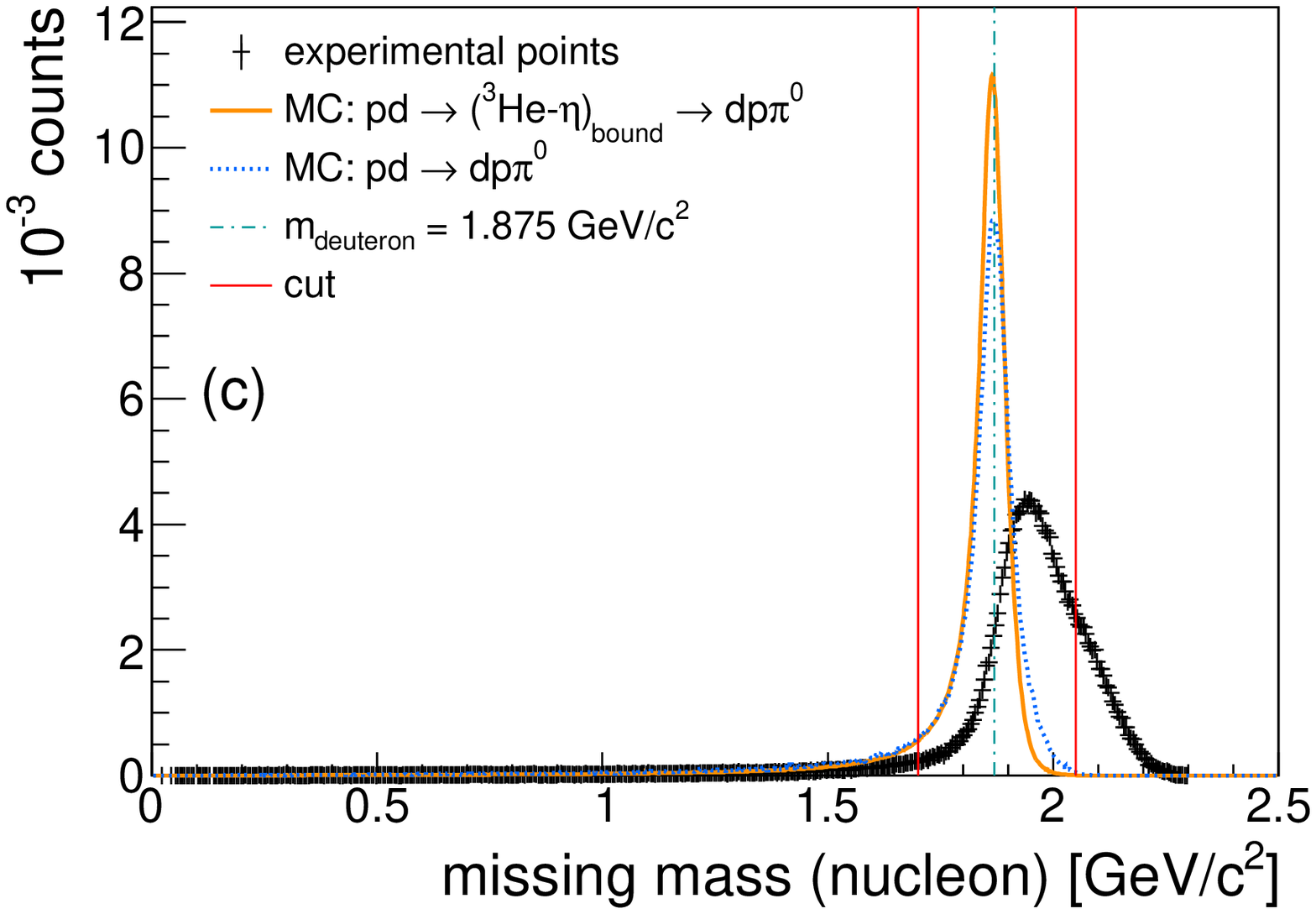}
	\includegraphics[width=0.99\columnwidth]{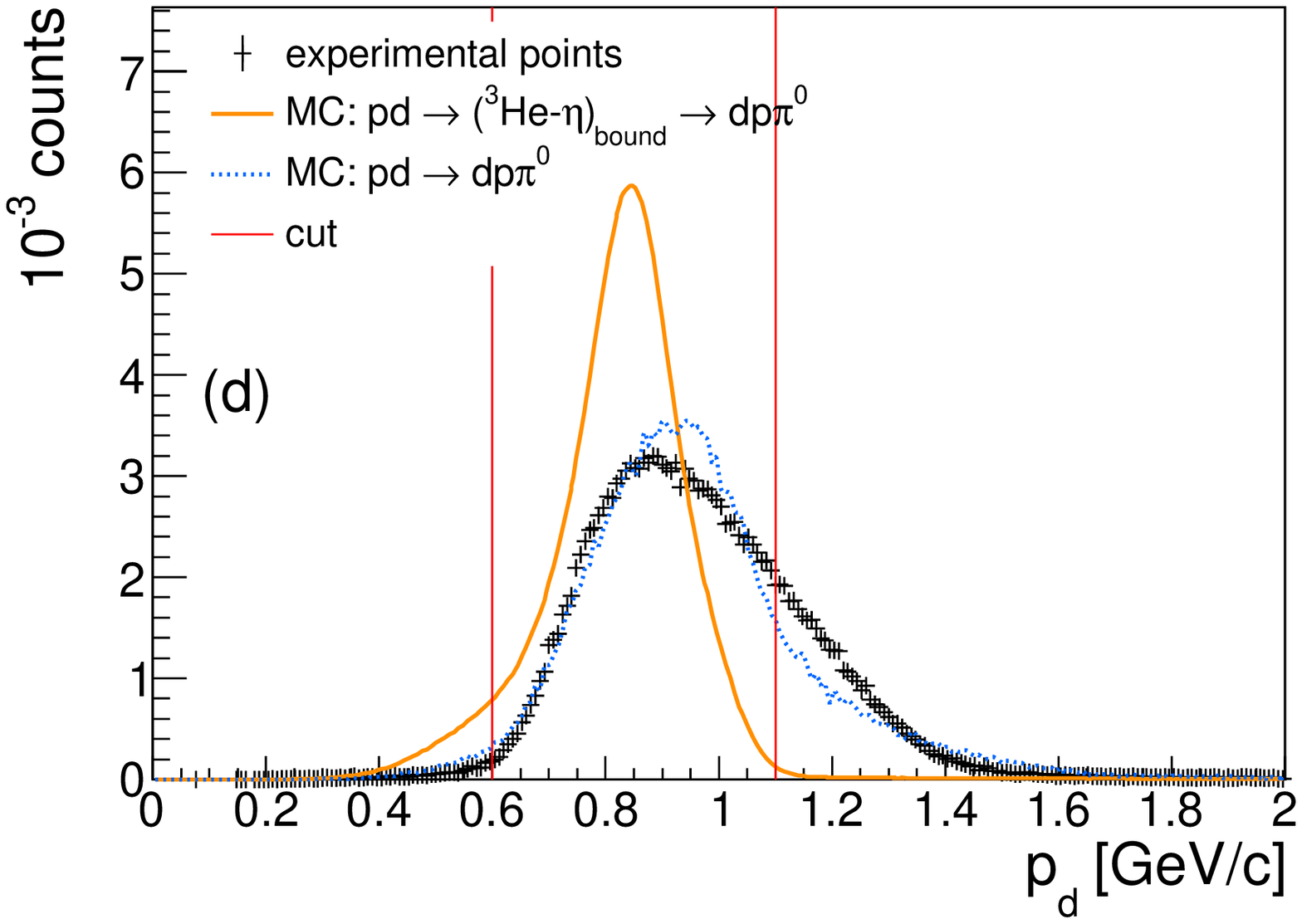}
	\caption{(Color online) (a) $\pi^{0}$ identification based on the two photon invariant mass spectrum, (b) $\pi^{0}$-proton opening angle in the c.m. frame $\vartheta^{c.m.}_{\pi^{0},p}$, (c) deuteron identification based on the missing mass technique, (d) the deuteron momentum distribution in the laboratory frame $p_{d}$. 
	Data are shown as black crosses. Orange solid and blue dotted curves show the simulation of signal and background from $pd \rightarrow d p \pi^{0}$ reaction respectively, while the red vertical lines indicate the boundary of the applied selection cuts.} 
	\label{sel_crit}
\end{figure*}

The neutral pions $\pi^{0}$ were identified on the basis of the invariant mass of two photons originating from their decays and measured in the SEC (Fig.~\ref{sel_crit}(a)). 

Deuterons which were not directly registered in the experiment were identified via the missing mass technique.
The events corresponding to $\eta$-mesic bound states were selected by applying cuts in the $\pi^{0}$-proton opening angle in the c.m. frame $\vartheta_{\pi^{0},p}^{c.m.}$, in the missing mass as well as in the deuteron momentum $p_{d}$ distributions. 
The spectra including experimental data and Monte-Carlo simulation for the signal and the dominant background $pd \rightarrow d p \pi^{0}$ process are presented in Fig.~\ref{sel_crit} with marked selection cuts. 

\begin{figure}[h!]
    \centering
	\includegraphics[width=0.99\columnwidth]{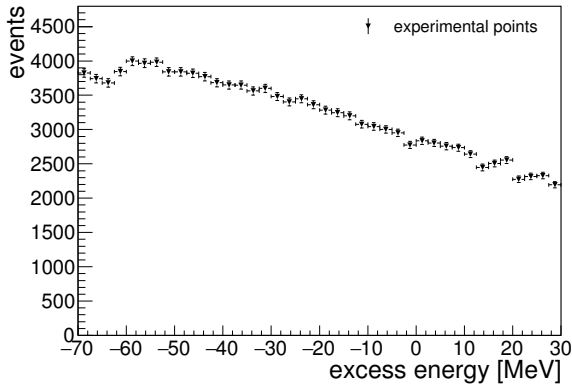}
	\caption{The number of selected events  for the $pd\rightarrow d p \pi^{0}$ reaction after application of all selection criteria.}
	\label{exc_fcn}
\end{figure}

The final number of selected events as a function of the excess energy Q for the $pd \rightarrow d p \pi^{0}$ reaction is shown in Fig.~\ref{exc_fcn}.
The excess energy range Q$\in(-70,30)$~MeV was divided into 40 intervals, each of width 2.5~MeV.

\newpage

\subsection{Luminosity and efficiency}

\begin{figure}[h!]
    \centering
	\includegraphics[width=0.99\columnwidth]{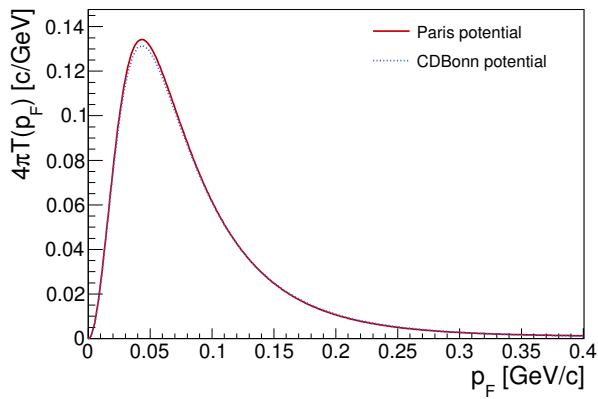}
	\caption{(Color online) Fermi momentum distribution of nucleons inside the deuteron for Paris (red solid line)~\cite{PARIS_model} and CDBonn (blue dotted line)~\cite{CDBONN_model} potential models.}
	\label{lum_calc}
\end{figure}

In order to determine the excitation function for the studied reaction the number of events in each excess energy interval has to be normalized by the integrated luminosity and corrected for the total efficiency. 
Since, during the beam ramping process the luminosity has varied due to the change of the beam-target overlap, the luminosity dependence on the excess energy L(Q) has been determined analysing the quasi-elastic proton-proton scattering process based on the method described in~\cite{czyzyk,Khreptak_epj}. 
For this purpose dedicated Monte Carlo simulation for $pd \rightarrow ppn_{spectator}$ reaction has been performed assuming that the beam protons scatter on the protons in the deuteron target and the neutrons from the deuteron play a role of spectators. 
The target nucleons momenta were generated isotropicaly with Fermi momentum distribution derived from the Paris ~\cite{PARIS_model} and the CDBonn potential models~\cite{CDBONN_model},
see Fig.~\ref{lum_calc}. 

In the analysis quasi-elastically scattered protons were searched for
with the primary events selection condition of exactly one charged particle in the Forward Detector and one charged particle in the Central Detector. Proton identification in the Central Detector was based on the selection criterium shown in Fig.~\ref{proton_cd}. 

\begin{figure*}[h!]
    \centering
    \includegraphics[width=0.99\columnwidth]{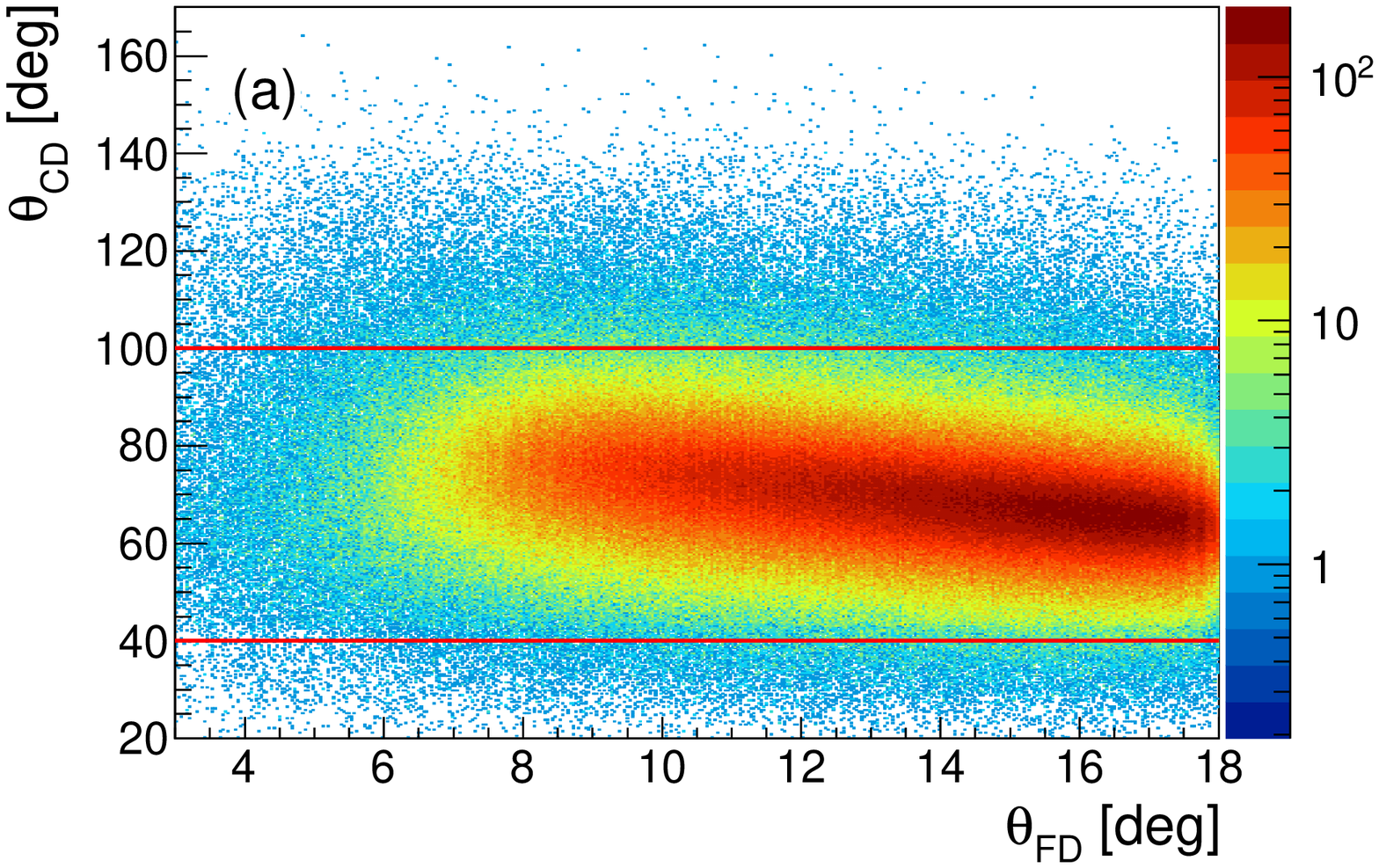} 
    \includegraphics[width=0.99\columnwidth]{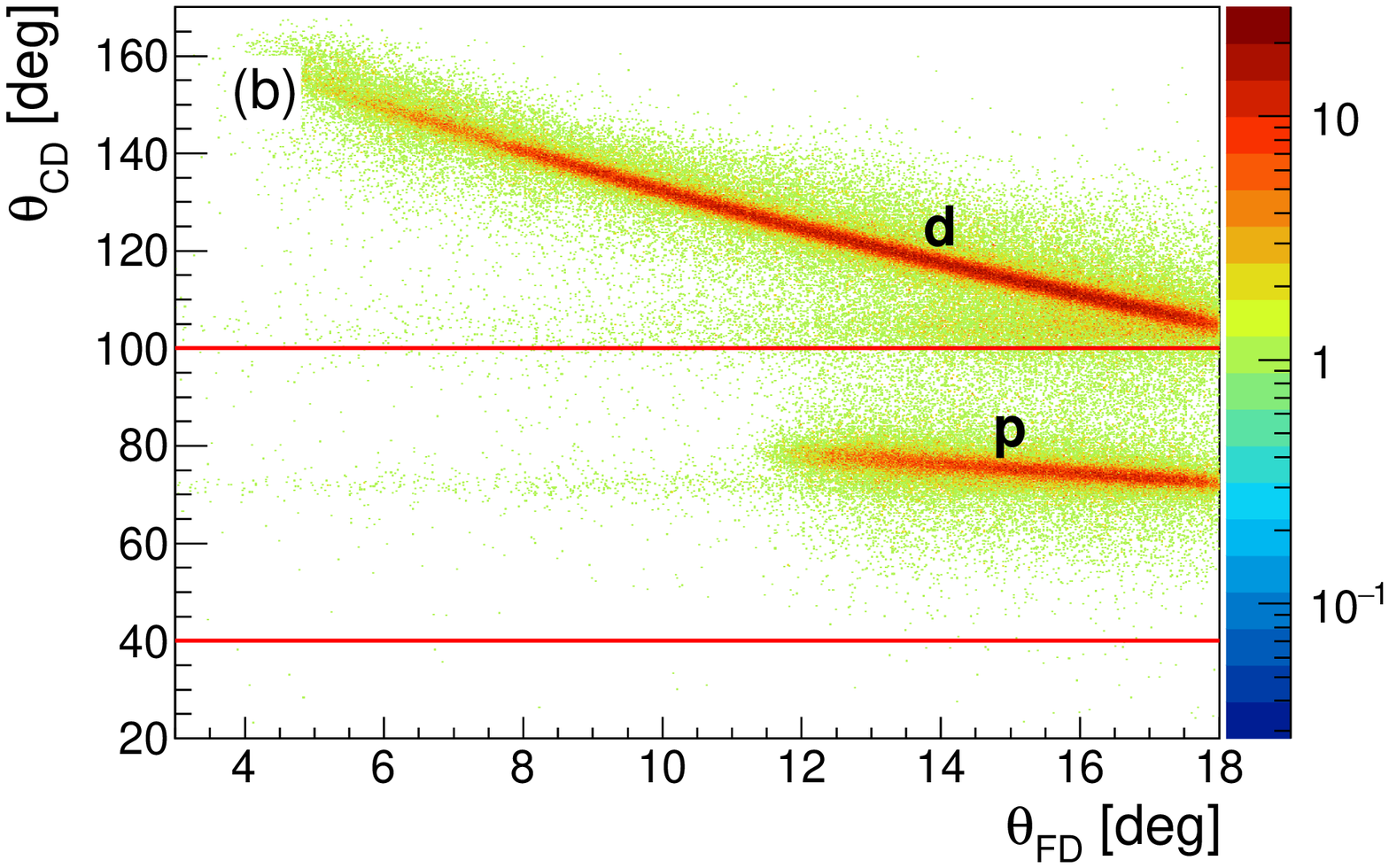} \\
    \includegraphics[width=0.99\columnwidth]{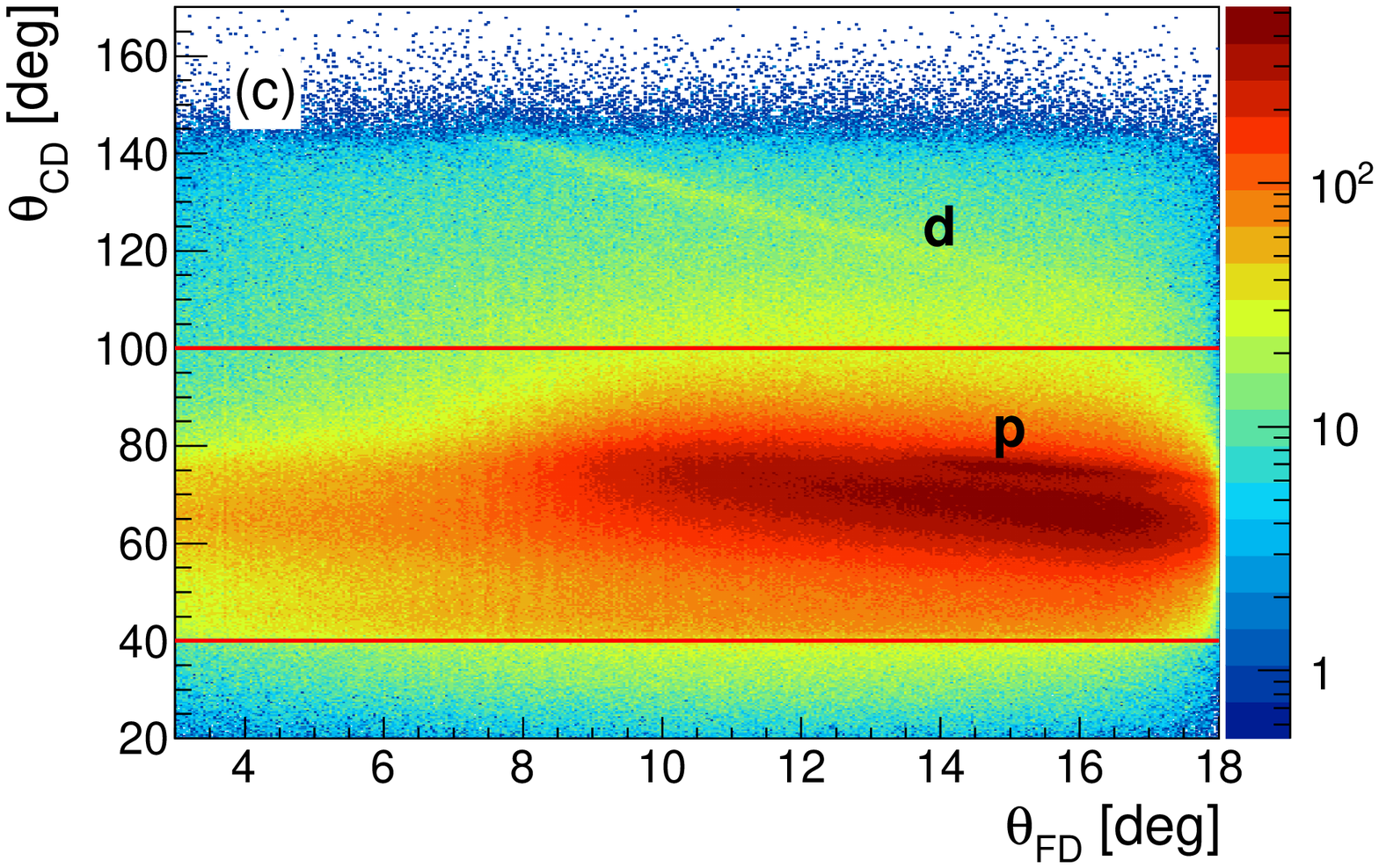}
    \includegraphics[width=0.99\columnwidth]{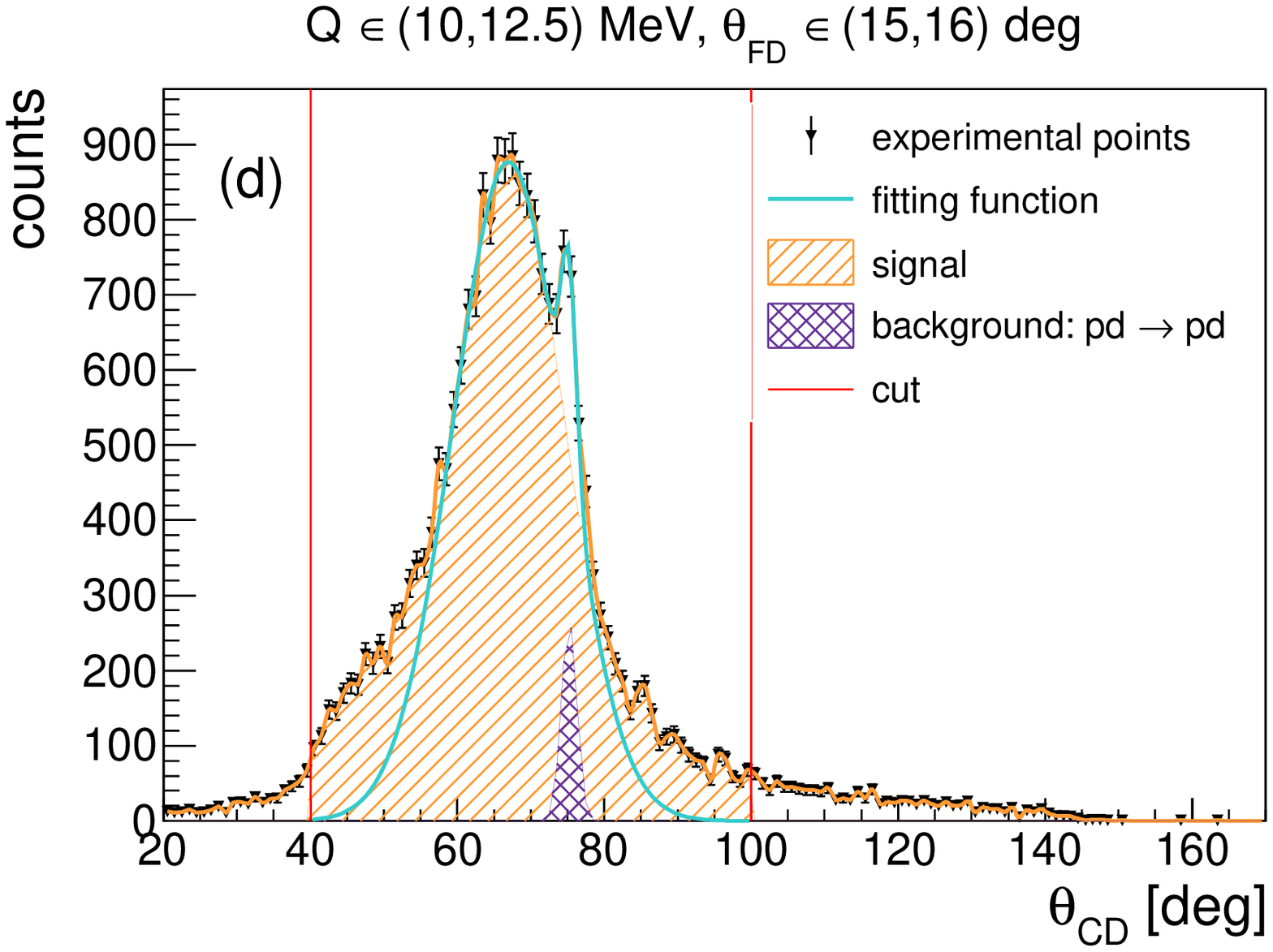}
    \caption{(Color online) Correlations between the polar angles of charged particles registered in the FD $\theta_{FD}$ and CD $\theta_{CD}$  obtained in the MC simulations for the $pd \rightarrow ppn_{spectator}$ (a) and $pd \rightarrow pd$ (b) reactions,  experimental data (c). 
    Note that the 2D spectra are in logarithmic scale. The applied cut is marked with red horizontal line. 
    The (d) panel shows an example of experimental distribution of $\theta_{CD}$ (black points), fitting function (cyan solid curve), signal from $pd \rightarrow ppn_{spectator}$ reaction (orange (light gray) area) and peak from background reaction $pd \rightarrow pd$ (purple checkered area) for Q~$\in (10,12.5)$~MeV and $\theta_{FD}\in(15,16)$~deg. 
    The applied cut ($\theta_{CD} \in(40,100)$~deg) is marked with red vertical lines.}
    \label{lumin_theta_FD_CD}
\end{figure*}

A part of the
background from elastic $pd \rightarrow pd$ scattering corresponding to deuterons was subtracted applying the criterium for polar angle $\theta_{CD} \in(40,100)$~deg, while part corresponding to protons was eliminated by fitting the $\theta_{CD}$ distribution for each interval of excess energy Q and polar angle $\theta_{FD}$ with the sum of two Gaussian functions (see Fig.~\ref{lumin_theta_FD_CD}(d)).
 
\begin{figure*}[h!]
	\centering
	\includegraphics[width=0.99\columnwidth]{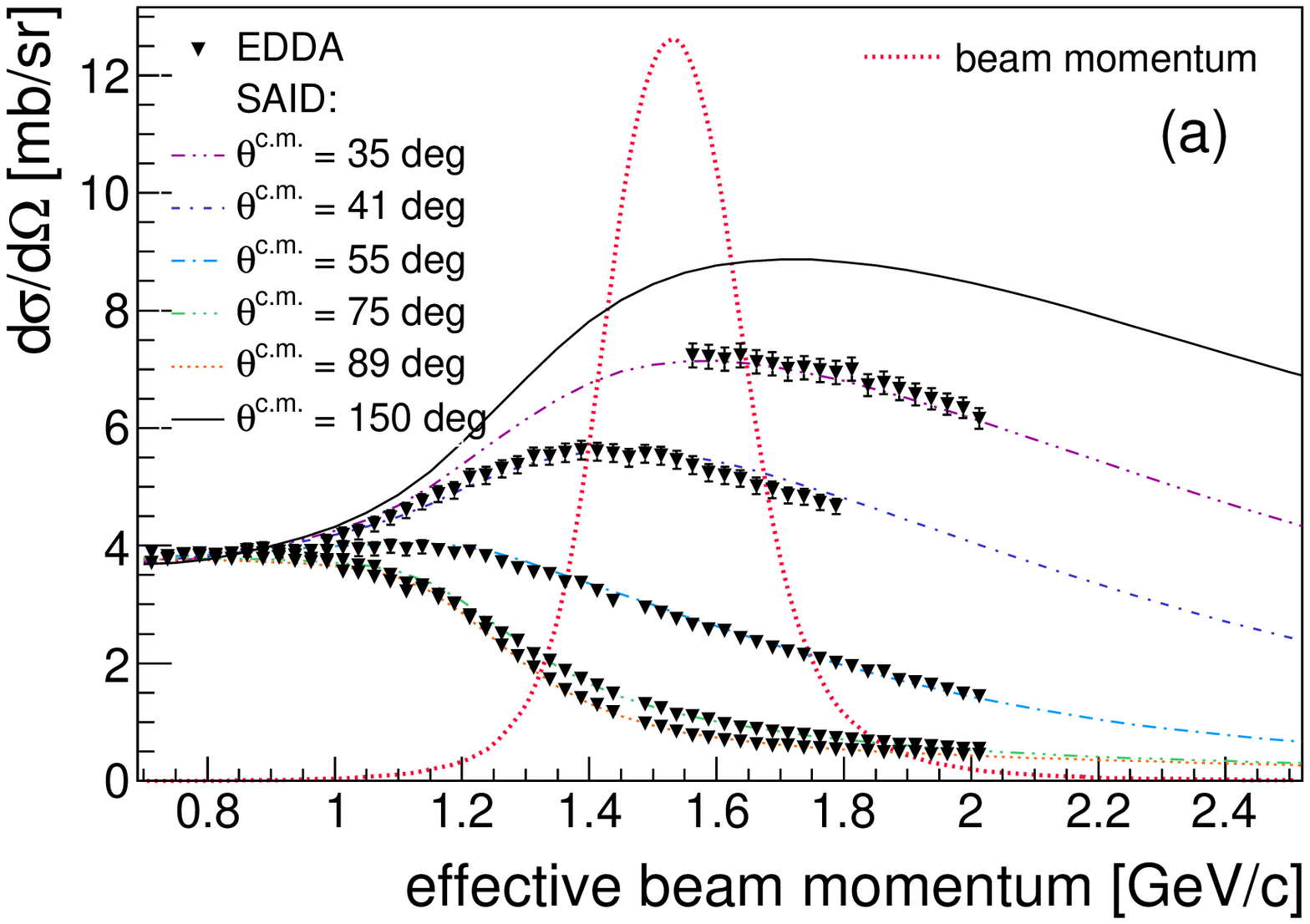}
	\includegraphics[width=0.99\columnwidth]{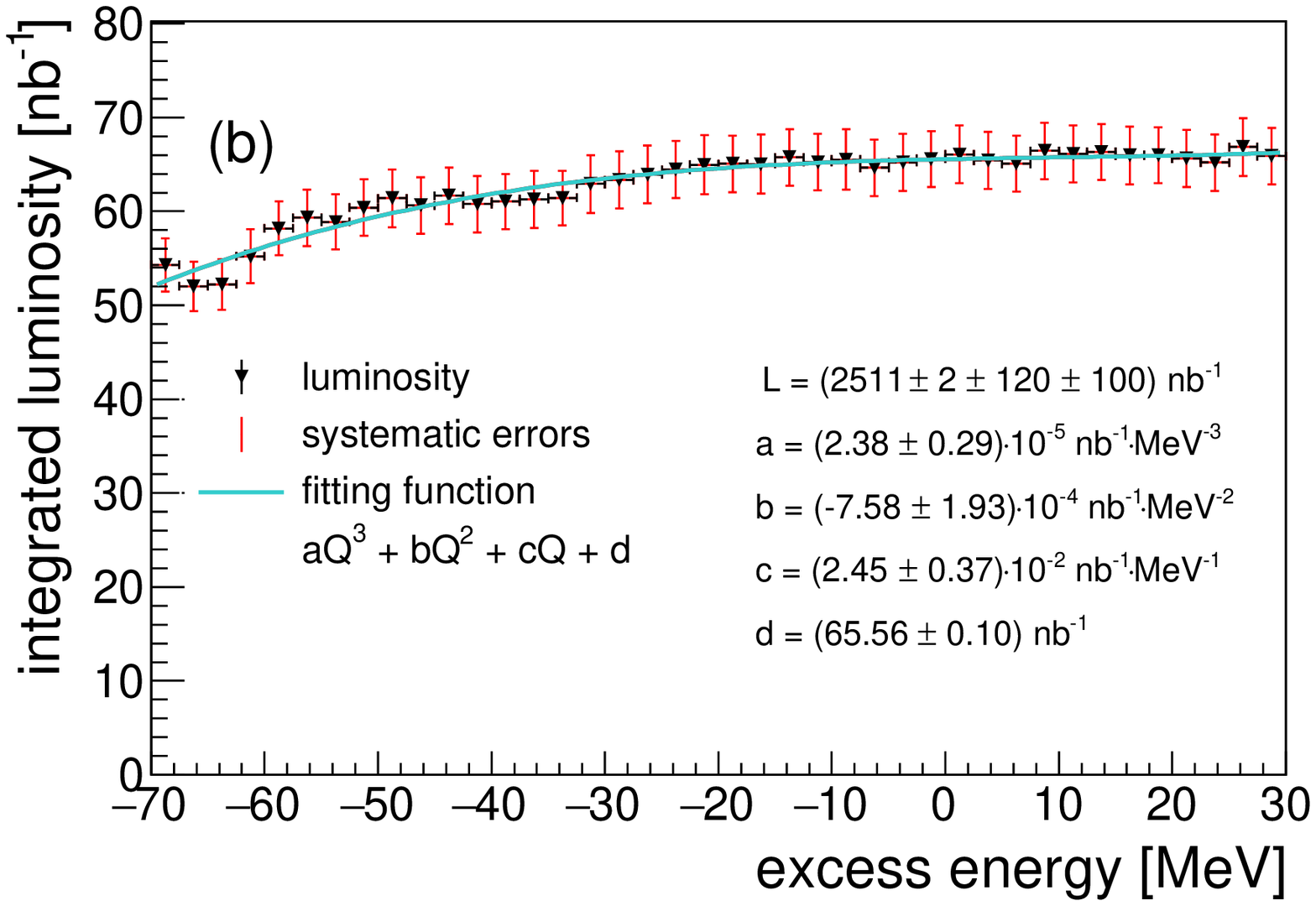}
	\caption{(Color online) (a) Differential cross sections for proton-proton elastic scattering as a function of the effective beam momentum for different values of the scattering angle $\theta^{c.m.}$ in the c.m. frame. Triangles show EDDA collaboration data~\cite{edda}. Curves denote SAID calculations~\cite{Arndt:2007PRC,PP_CS_link}. The pink dotted line presents the distribution of the effective beam momentum obtained from simulations.
	(b) Integrated luminosity calculated based on experimental data for quasifree $pd \rightarrow ppn_{spectator}$ reaction with statistical (black points) and systematic (red vertical bars) errors fitted with third degree polynomial function (cyan curve).}
	\label{XS_pp_lum} 
\end{figure*}

In order to determine the integrated luminosity the number of reconstructed events obtained from Monte Carlo simulation was weighted with the values of the differential cross section for the quasi-free proton-proton scattering, which is uniquely determined by the scattering angle and the total proton-proton collision energy. For the estimation of the differential cross-sections the data for elastic proton-proton scattering~\cite{Arndt:2007PRC,PP_CS_link,edda} has been used 
(see Fig.~\ref{XS_pp_lum}(a)). 
The integrated luminosity dependence on the excess energy is presented 
in Fig.~\ref{XS_pp_lum}(b) 
and its total value is equal to $2511 \pm 2(stat.) \pm 120(syst.) \pm 100(norm.)$ $\textrm{nb}^{-1}$, where the statistical, systematic and normalization errors are indicated, respectively. 
In the calculations the shadowing effect equals 4.5\%~\cite{shading_effect} caused by the neutron shading the scattered protons. 
The total integrated luminosity is consistent within systematic and normalization errors with the luminosity determined for the current experiment based on two alternative methods presented in Refs.~\cite{AdlarsonPLB2020,Rundel_PhD}.

\begin{figure}[h!]
    \centering
	\includegraphics[width=0.99\columnwidth]{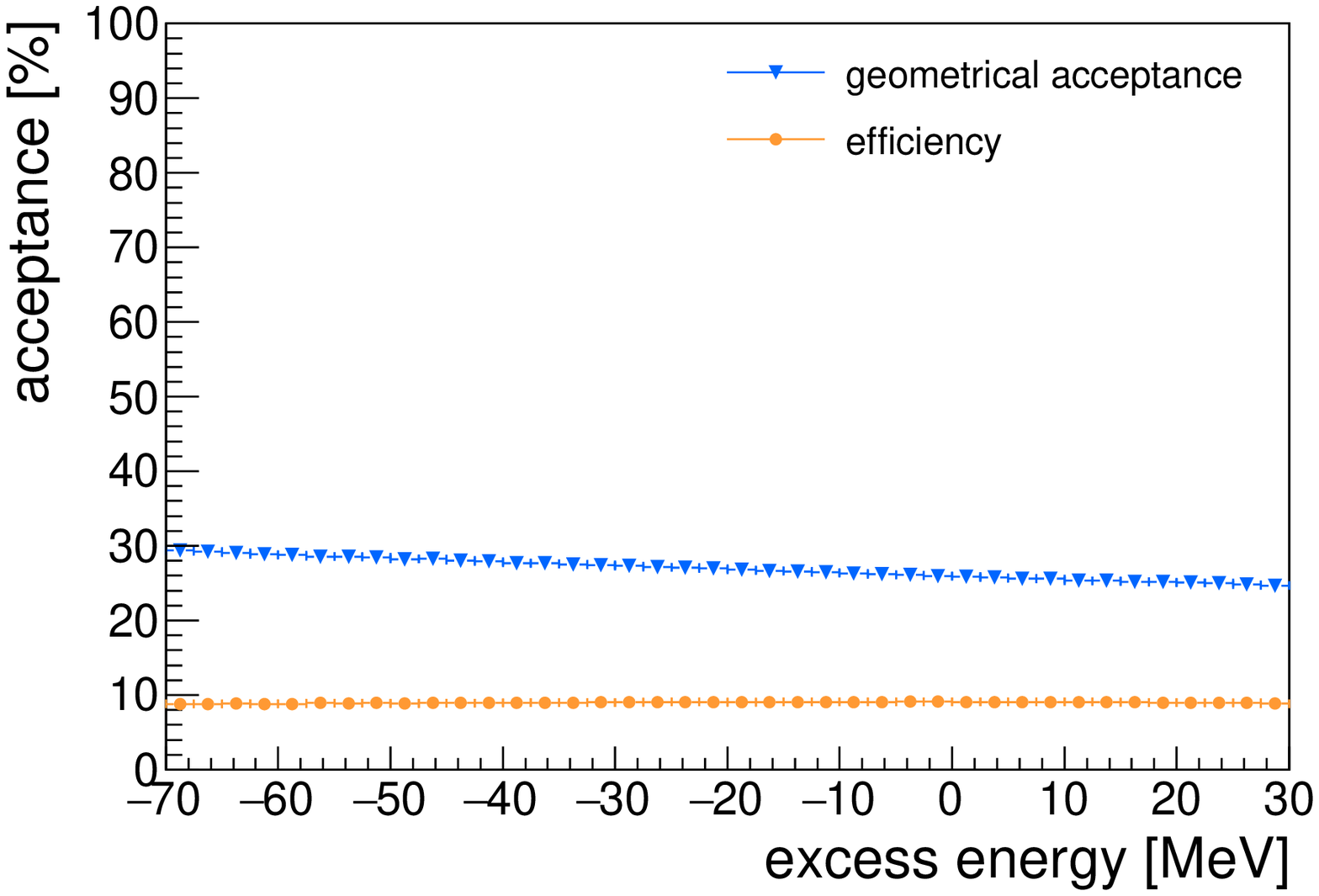}
	\vspace{-0.5cm}
	\caption{(Color online) Geometrical acceptance (blue triangles) and efficiency (orange circles) for the 
	$pd\rightarrow (^{3}\hspace{-0.03cm}\textrm{He-}\eta)_{bound} \rightarrow d p \pi^{0}$ reaction as a function of excess energy.}
	\label{eff}
\end{figure}

The Monte Carlo simulations for the $pd \rightarrow (^{3}\hspace{-0.03cm}\textrm{He-}\eta)_{bound} \rightarrow d p \pi^{0}$ process allowed one to determine detection and reconstruction efficiency as a function of the excess energy Q. The obtained geometrical acceptance is equal to about 30\% while the full efficiency including all applied selection criteria is about 9\% (see Fig.~\ref{eff}).

\subsection{Upper limit of the total cross section}

\begin{figure}[h!]
    \centering
	\includegraphics[width=0.99\columnwidth]{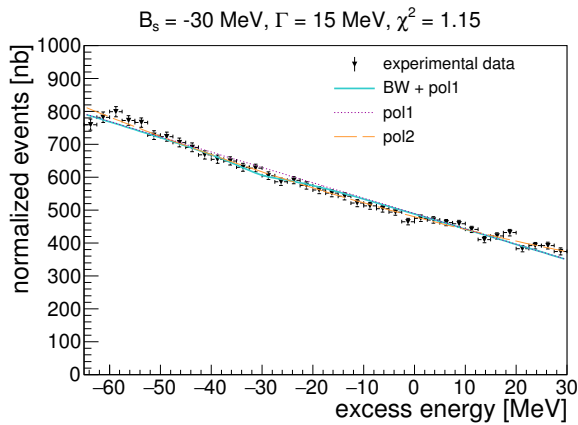}
	\caption{(Color online) Experimental excitation function for the 
	$pd \rightarrow d p \pi^{0}$ process obtained after applying the selection criteria described in the text, correction by the efficiency, and normalization by the corresponding integrated luminosity. The cyan solid line represents a fit with a first order polynomial combined with a Breit-Wigner function with fixed binding energy and width equal to $-30$~MeV and 15~MeV, respectively. The purple dotted and orange dashed lines show the first and second order polynomial (describing the background), respectively.}
	\label{fit}
\end{figure}

The final excitation function (Fig.~\ref{fit}) was obtained by correcting the number of events identified as $pd \rightarrow (^{3}\hspace{-0.03cm}\textrm{He-}\eta)_{bound} \rightarrow d p \pi^{0}$ for the efficiency (Fig.~\ref{eff}) and normalizing by the luminosity 
(Fig.~\ref{XS_pp_lum}(b)). 
The excitation curve does not show any structure that could be interpreted as an indication for the $\eta$-mesic $^{3}\hspace{-0.03cm}\textrm{He}$. 

Hence, the upper limit of the total cross-section for the $^{3}\hspace{-0.03cm}\textrm{He-}\eta$ bound state production and its decay to $d p \pi^{0}$ channel was evaluated. 
In order to quantitatively estimate the upper limit, a fit to the excitation function with a polynomial describing the background (first and second order) combined with a Breit-Wigner function (for the signal) was performed.
In the fit the polynomial coefficients and the normalization of the Breit-Wigner amplitude were treated as free parameters, while the binding energy $B_{s}$ and the width $\Gamma$ were fixed in the range from $-40$~MeV to 0~MeV and from 5~MeV to 50~MeV, respectively. 
An example excitation function with the fit result for binding energy $-30$~MeV and width 15~MeV is presented in Fig.~\ref{fit}. 

The upper limit of the total cross section was determined based
on the uncertainty of the amplitude obtained from the fit $\Delta \sigma_{A}$:
\begin{equation}
    \sigma_{upper}^{CL=90\%}(B_{s},\Gamma) = k \cdot \Delta\sigma_{A},  
\end{equation}

\noindent
where $k$ is the statistical factor equal to 1.64 corresponding to 90\% confidence level (CL) as given by the Particle
Data Group, PDG~\cite{PDG}. 

\begin{figure}[h!]
    \centering
	\includegraphics[width=0.99\columnwidth]{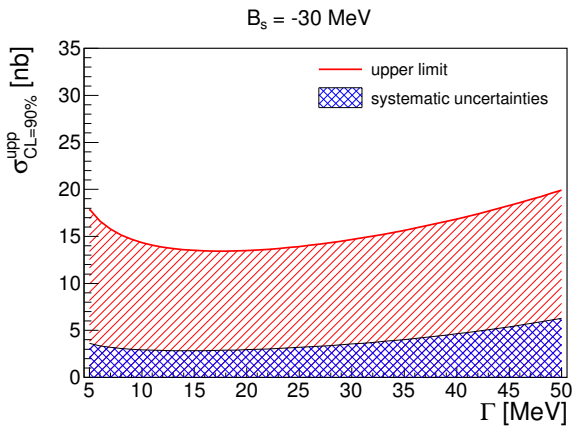}
	\caption{(Color online) The upper limit at the 90\% confidence level of the total cross section for formation of the $^{3}\hspace{-0.03cm}\textrm{He-}\eta$ bound state and its decay via the $pd\rightarrow (^{3}\hspace{-0.03cm}\textrm{He-}\eta)_{bound} \rightarrow d p \pi^{0}$ reaction as a function of the width of the bound state. The binding energy was fixed to $B_{s} = -30$~MeV. 
	The blue checkered area at the bottom represents the systematic uncertainties.}
	\label{upp_lim_30}
\end{figure}

\begin{figure}[h!]
    \centering
	\includegraphics[width=0.99\columnwidth]{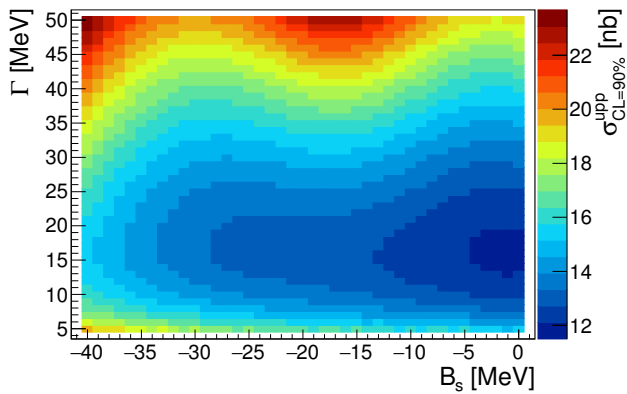}
	\caption{(Color online) The upper limit of the total cross section at the 90\% confidence level obtained based on excitation curves fit assuming different bound state parameters, $B_{s}$ and $\Gamma$.}
	\label{upp_lim_all}
\end{figure}

The upper limit obtained by averaging the results derived from fits with a background described by the linear and quadratic functions for different values of $B_{s}$ and $\Gamma$ is presented in Table~\ref{tab:upp_limit}. 
It varies between 13 to 24 nb and depends mainly on the width of the bound state while is not sensitive to the binding energy. 
The result for $B_{s} = -30$~MeV is shown in Fig.~\ref{upp_lim_30}. 
The blue checkered area denotes the systematic errors described in the next section.
The obtained upper limit as a function of $B_s$ and $\Gamma$ is presented in Fig.~\ref{upp_lim_all}.

\begin{table}[h]
  \begin{center}
    \caption{The upper limit for the cross section for the bound state formation and decay in the $pd\rightarrow (^{3}\hspace{-0.03cm}\textrm{He-}\eta)_{bound} \rightarrow d p \pi^{0}$ process, determined at the 90\% confidence level. 
    The values were obtained by fitting excitation curve with a Breit-Wigner function combined with the first and second order polynomial with different fixed bound state parameters, $B_{s}$ and $\Gamma$.}
    \label{tab:upp_limit}
    \begin{ruledtabular}
    \begin{tabular}{c c c c c c}      
      $B_{s}$ & $\Gamma$ & $\sigma_{upper}^{CL=90\%}$ & $B_{s}$ & $\Gamma$ & $\sigma_{upper}^{CL=90\%}$ \\
      
      [MeV] & [MeV] & [nb] & [MeV] & [MeV] & [nb] \\
      \hline  
      -40   & 5   & 19.74  & -20   & 5     & 16.85 \\
      -40   & 10  & 16.08  & -20   & 10    & 13.64 \\
      -40   & 20  & 15.61  & -20   & 20    & 13.19 \\
      -40   & 30  & 17.35  & -20   & 30    & 14.86 \\
      -40   & 40  & 20.14  & -20   & 40    & 17.86 \\
      -40   & 50  & 23.67  & -20   & 50    & 22.21 \\
      -30   & 5   & 17.91  & -10   & 5     & 16.11 \\
      -30   & 10  & 14.34  & -10   & 10    & 13.07 \\
      -30   & 20  & 13.49  & -10   & 20    & 12.67 \\
      -30   & 30  & 14.66  & -10   & 30    & 14.23 \\
      -30   & 40  & 16.85  & -10   & 40    & 16.96 \\
      -30   & 50  & 19.92  & -10   & 50    & 20.79 \\      
    \end{tabular}
	\end{ruledtabular}
  \end{center}
\end{table}

\subsection{Systematics}

Systematic checks were performed just as in the previous analyses presented in Refs.~\cite{AdlarsonPRC2013,Adlarson:2016dme}. The upper limit of the total cross section obtained in the 
$pd \rightarrow (^{3}\hspace{-0.03cm}\textrm{He-}\eta)_{bound} \rightarrow d p \pi^{0}$ 
reaction analysis is sensitive to the variation of the selection criteria, systematic error of the luminosity determination, and application of different theoretical models. 

Changing the selection criteria applied in analysis within $\pm10\%$ results in the systematic error of about 8.5\%. 

Overall systematic and normalization errors of the luminosity determined based on the quasi-free $pp$ reaction are equal to 4.8\% and 4\%, respectively, and are another contribution to the systematic uncertainty of the upper limit. 

The description of the background with quadratic and linear functions introduces additional systematic uncertainty, which is estimated as
\begin{equation}
    \delta = \frac{\sigma_{quad} - \sigma_{lin}}{2}.
\end{equation}
This systematic error changes from about 2\% (for $\Gamma = 5$~MeV) to 24\% ($\Gamma = 50$~MeV).

An important source of systematic errors comes from the assumption of the $\textrm{N}^{\ast}$ momentum distribution inside the $^{3}\hspace{-0.03cm}\textrm{He}$ nucleus applied in the simulations. 
The current analysis was performed with the Fermi momentum distribution for $\textrm{N}^{\ast}$ determined for binding energy $-0.53$~MeV by Kelkar et al.~\cite{Kelkar:2019hjm,Kelkar:2020wmh}. 
In addition, in this analysis the simulations were also performed assuming that the $\textrm{N}^{\ast}$ resonance in the c.m. frame moves with a momentum distribution similar to that of protons inside $^{3}\hspace{-0.03cm}\textrm{He}$~\cite{Nogga:PRC2003} (see the blue dotted line in Fig.~\ref{distr}). 
The choice of the alternative model does not influence the experimental method but it affects the acceptance of the deuterons in the FD, which is connected with the fact that the momentum distribution of protons inside $^{3}\hspace{-0.03cm}\textrm{He}$ is peaked at higher value with respect to the $\textrm{N}^{\ast}$ distribution in the $\textrm{N}^{\ast}\textrm{-}d$ system. 
It provides a systematic error of about 17\%.

Adding the above-estimated contributions in quadrature we obtain systematic uncertainty of the upper limit that varies from 20\% to 31\%. 
The systematic uncertainties are presented by the blue checkered area in Fig.~\ref{upp_lim_30}.

\section{Conclusion}

In order to search for evidence of a possible
$^{3}\hspace{-0.03cm}\textrm{He}\eta$ bound state we performed measurements of the proton beam scattering on a deuteron target with the WASA-at-COSY detector.
The analysis was based on the determination of the excitation function for the $pd \rightarrow d p \pi^{0}$ process.
The applied selection criteria were inferred from Monte-Carlo simulations based on the assumption that the $\textrm{N}^{\ast}$ resonance momentum in the $\textrm{N}^{\ast}$-deuteron bound state is distributed according to the recent theoretical modelling in~\cite{Kelkar:2019hjm,Kelkar:2020wmh}. 

Narrow resonance-like structure associated with an $\eta$-mesic $^{3}\hspace{-0.03cm}\textrm{He}$ bound state was not observed. 
Therefore, the upper limit for the total cross sections for the 
$pd \rightarrow (^{3}\hspace{-0.03cm}\textrm{He-}\eta)_{bound} \rightarrow d p \pi^{0}$ process was estimated and varies from 13 to 24 nb depending on the bound state parameters $B_{s} \in (0,40)$~MeV and $\Gamma \in (0,50)$~MeV.

The upper limit obtained in this analysis for the 
$pd \rightarrow (^{3}\hspace{-0.03cm}\textrm{He-}\eta)_{bound} \rightarrow d p \pi^{0}$ 
reaction is about 3 times lower than the limit of 70~nb~\cite{Moskal:2010ee,Smyrski:2007NPA} determined by the COSY-11 collaboration for the 
$pd \rightarrow (^{3}\hspace{-0.03cm}\textrm{He-}\eta)_{bound} \rightarrow {^{3}\hspace{-0.03cm}\textrm{He}} \pi^{0}$ process.   
The limit about 24~nb found here compares with the total cross section for $\eta$ meson production above threshold in $dp$ collisions which is about 400~nb~\cite{SmyrskiPLB2007}. 
In $dd$ collisions the limits obtained by the WASA-at-COSY Collaboration for the $dd \rightarrow {^{3}\hspace{-0.03cm}\textrm{He}} n \pi^{0}$, $dd \rightarrow {^{3}\hspace{-0.03cm}\textrm{He}} p \pi^{-}$ processes ($2.5-7$ nb)~\cite{AdlarsonPRC2013,Adlarson:2016dme} compare with the total cross section
15~nb~\cite{Willis:1997PLB} for $\eta$ production above threshold.
These measurements provide an important constraint for models of He-$\eta$ bound state production. 
Within the limits determined here, bound states predicted with $\eta$-nucleon
scattering lengths about 1~fm remain a possibility.

\begin{acknowledgments}
We acknowledge the support from the Polish National Science Center through grant No. 2016/23/B/ST2/00784. Theoretical parts of this work was partly supported by the Faculty of Science, Universidad de los Andes, Colombia, through project number P18.160322.001-17, and by JSPS KAKENHI Grant Numbers JP16K05355 (S.H.) in Japan.
\end{acknowledgments}


\bibliography{mybibfile}


\bibliographystyle{unsrt}

\end{document}